 \documentclass[final,5p,times]{elsarticle}

\usepackage{amsfonts}
\usepackage{amsmath}

\usepackage{mathabx}						
\usepackage{subcaption,caption}
\usepackage{tikz,pgfplots}
\usepackage{epsfig}
\usepackage{mathtools}
\usepackage{booktabs}                       
\usepackage{tabularx}                       
\graphicspath{{FIGURES/}}                   
\usepackage{accents}                        
\usepackage{supertabular}
\usepackage{multirow}
\usepackage{longtable}
\usepackage{pdflscape}
\usepackage{afterpage}
\usepackage{placeins}
\usepackage{capt-of}
\usepackage{comment}                        
\usepackage{bm}

    \makeatletter

\newsavebox{\bigleftbox}

\newcounter{subassumption}[asu]

\makeatletter
\renewcommand{\p@subassumption}{\theasu}
\makeatother

\makeatletter
\def\BState{\State\hskip-\ALG@thistlm}
\makeatother

\usepackage[colorlinks=true,urlcolor=blue]{hyperref} 

\journal{Energy.}

\usepackage{algorithm}
\usepackage{algpseudocode}

\usepackage{framed} 
\usepackage{multicol} 
\usepackage{nomencl} 
\makenomenclature
\renewcommand*\nompreamble{\begin{multicols}{2}}
\renewcommand*\nompostamble{\end{multicols}}

\usepackage{etoolbox}
\renewcommand\nomgroup[1]{%
  \item[\bfseries
  \ifstrequal{#1}{G}{Greek symbols}{%
  \ifstrequal{#1}{S}{Subscripts}{%
  \ifstrequal{#1}{U}{Superscripts}{}}}%
]}

\begin{document}

\begin{frontmatter}


\title{
A 1D Model for the Unsteady Gas Dynamics of Ejectors}


\author[vki,ucl]{Jan Van den Berghe\corref{cor1}}
\ead{jan.van.den.berghe@vki.ac.be}
\author[vki]{Bruno R. B. Dias}
\author[ucl]{Yann Bartosiewicz}
\author[vki]{Miguel A. Mendez}

\cortext[cor1]{Corresponding author}
\address[vki]{von Karman Institute for Fluid Dynamics, Waterloosesteenweg 72, 1640 Sint-Genesius-Rode, Belgium}
\address[ucl]{Université Catholique de Louvain. Place de l'Université 1, 1348 Louvain-la-Neuve, Belgium}


\begin{abstract}
We propose a 1D unsteady model for supersonic single-phase ejectors. The model treats an ejector as a pipe network with two inputs and one output and combines a 1D gas dynamics formulation in each `pipe' with a junction model for entrainment and mixing. The model is calibrated and validated on experimental data in steady-state conditions and used to analyze the choking mechanism for the mixed flow. The model was then benchmarked against 2D URANS simulations to predict the ejector response to a sudden change in operating conditions, producing traveling waves. The results show that the model can correctly predict the ejector performance and the stream-wise evolution of relevant integral quantities (e.g. mass flow rates and momentum) in both steady and transient conditions.
\end{abstract}


\end{frontmatter}

\begin{table*}[!t]   

\begin{framed}
\nomenclature[]{$a$}{speed of sound [m/s]}
\nomenclature[]{$c$}{constant in Blasius' formula [-]}
\nomenclature[]{$\bm{c}$}{vector of operating conditions $\left[p_{t,p}, T_{t,p}, p_{t,s}, T_{t,s}, p_d\right]^T$}
\nomenclature[]{$e$}{internal energy [J/kg]}
\nomenclature[]{$f$}{friction coefficient [-]}
\nomenclature[]{$h$}{enthalpy [J/kg]}
\nomenclature[]{$k$}{turbulent kinetic energy [J/kg]}
\nomenclature[]{$l_{mix}$}{turbulent mixing length [m]}
\nomenclature[]{$\dot{m}$}{mass flow rate [kg/s]}
\nomenclature[]{$\bm{n}$}{normal vector [-]}
\nomenclature[]{$p$}{pressure [Pa]}
\nomenclature[]{$P$}{cross-sectional perimeter [m]}
\nomenclature[]{$s$}{entropy [J/kg/K]}
\nomenclature[]{$t$}{time [s]}
\nomenclature[]{$u$}{$x$-component of velocity [m/s]}
\nomenclature[]{$\bm{v}$}{velocity vector [m/s]}
\nomenclature[]{$\bm{w}$}{vector of model coefficients $\left[\eta_m, c_p, c_s, c_m \right]^T$ [-]}
\nomenclature[]{$x$}{$x$-coordinate [m]}
\nomenclature[]{$A$}{cross-sectional area [m²]}
\nomenclature[]{$\bm{A}$}{vector of ejector geometry $[A_p, A_s, A_m]^T$ [m²]}
\nomenclature[]{$C_v$}{constant volume heat capacity [J/kg/K]}
\nomenclature[]{$\bm{F}$}{flux vector [-]}
\nomenclature[]{$L$}{length of a domain [m]}
\nomenclature[]{$M$}{Mach number [-]}
\nomenclature[]{$\mathcal{M}$}{momentum flow rate [N]}
\nomenclature[]{$R$}{gas constant [J/kg/K]}
\nomenclature[]{$Re$}{Reynolds number [-]}
\nomenclature[]{$\bm{S}$}{source vector [-]}
\nomenclature[]{$T$}{temperature [K]}
\nomenclature[]{$TI$}{turbulence intensity [-]}
\nomenclature[]{$\bm{U}$}{state vector [-]}

\nomenclature[G]{$\gamma$}{ratio of specific heat capacities [-]} 
\nomenclature[G]{$\Delta t$}{time step [s]} 
\nomenclature[G]{$\eta_m$}{mixing efficiency [-]}
\nomenclature[G]{$\rho$}{density [kg/m³]}
\nomenclature[G]{$\phi$}{entrainment ratio [-]}
\nomenclature[G]{$\omega$}{specific dissipation [$\text{s}^{-1}$]}

\nomenclature[S]{$d$}{diffuser exit}
\nomenclature[S]{$m$}{mixing}
\nomenclature[S]{$p$}{primary}
\nomenclature[S]{$s$}{secondary}
\nomenclature[S]{$t$}{total quantity (vs. static)}
\nomenclature[S]{$th$}{throat of the primary nozzle}
\nomenclature[S]{$L$}{left star state}
\nomenclature[S]{$R$}{right star state}

\nomenclature[U]{$\bullet^*$}{star state ($\bm{U}$)}
\nomenclature[U]{$\bullet^*$}{critical operating point ($p$, $\phi$)}
\nomenclature[U]{$\bar{\bullet}$}{state in the internal domain}
\nomenclature[U]{$\bullet^0$}{breakdown point}

\printnomenclature

\end{framed}

\end{table*}

\section{Introduction}

Ejectors are passive devices for mixing, compression or aspiration. Their operating principle is based on expanding a high-pressure flow into a mixing chamber such that shear and pressure effects entrain a secondary flow. Having no moving parts, ejectors need little maintenance and have low operational costs, no restriction on the working fluids, and a long service life \cite{grazzini_book}. They can be used to decrease the electricity-driven compression work and reduce throttling losses in refrigeration systems \cite{BESAGNI2016373}, recover low-grade waste heat \cite{BESAGNI2019998}, minimize jet noise and increase thrust in an aircraft \cite{hunter2002} and more \cite{aidoun2019}.

The pre-design of an ejector (or its integration into complex systems) is traditionally carried out using Lumped Parametric Models (LPM). These were pioneered by Keenan \emph{et al} in 1942 \cite{keenan1942} and are based on integral balances of mass, momentum and energy over key control volumes (see \cite{aidoun2019} for a review). More recent variants of these models include the formulation of empirical or semi-empirical laws to model losses \cite{eames1995}, the shock patterns during the mixing \cite{zhu2007shock} and the sub-critical operation \cite{chen2013}. Depending on the operating conditions, the mixing modeling can be described according to the `Fabri-choking' theory, in which the entrained flow is choked \cite{fabri1958, mundaybagster1977} or the `compound-choking' theory, in which the mixed flow is choked \cite{LAMBERTS2018_compound, CROQUER2021120396,METSUE2021121856}.

While these models are fundamental design tools because of their low computational cost, their main limitations are (1) the need for closure parameters (e.g. isoentropic or mixing efficiencies, see \cite{BESAGNI2019998, tashtoush2019comprehensive}), which require calibration on numerical or experimental data \cite{liu2014} and (2) their zero-dimensional formulation, which does not allow to study the impact of the nozzle position, mixing chamber contours or downstream evolution of velocity, pressure or Mach number. Moreover, these models are stationary, hence unable to describe transient phenomena such as traveling waves after a sudden transition between operating conditions. 

Computational Fluid Dynamics (CFD) has emerged as a valid alternative design tool, allowing for a detailed description of the turbulent mixing and complex flow features such as oblique shock patterns and shock-boundary layer interaction. Numerical studies have focused on RANS turbulence modelling \cite{BARTOSIEWICZ200556} or the definition of macroscopic variables to quantify the mass, momentum and energy transfers between the two streams \cite{LAMBERTS201723}. Supported by surrogate models to lower computational costs, CFD simulations of ejectors have been used to drive design optimization in single-phase configuration in \cite{EXPOSITOCARRILLO201846, sierra2018shape, ringstad2021machine}.

Nevertheless, the high computational cost of CFD simulation motivates the interest in simplified models that bridge the gap between the 0D formulations in LPM models and the 2D/3D formulations in classic CFD. Notable efforts in this direction are the 1D formulation proposed by Banasiak and Hafner \cite{BANASIAK20112235}, who derive a 1D boundary value problem embedding the conservation laws in two parallel domains, and the quasi-2D approach by del Valle et al. \cite{delvalle2012}, which is based on potential flow theory for the core of the primary flow and Prandtl-Meyer expansion theory for the dividing streamline between the streams.

The aforementioned approaches provide a trade-off between computational cost and detailed local information but rely on the steady-state assumption. Several applications feature ejectors in transient conditions: these include pulsed solar refrigeration \cite{HUANG2006476}, pulsed thrust augmentation \cite{binder1975improvement, choutapalli2012, quinn1977, wilson2003, paxson2002unsteady}, pulsed mixing for combustion or compactness \cite{parikh1982, viets1981unsteady} and pulse detonation engines \cite{allgood2008performance}. The time scale in these applications ranges from a few seconds to milliseconds, either leading to a sequence of quasi-steady states or to travelling pressure waves \cite{parikh1982, viets1981unsteady, abdel1988duct, rao2014observations}. These propagate mostly in the streamwise direction due to the slender geometry of ejectors. Such flow fields are often encountered in pipeline networks, where coupling 1D domains for unsteady gas dynamics has been the subject of several studies \cite{reigstad2014, morin2015, lang}. 


This work proposes a 1D gas dynamic approach to the unsteady modeling of ejectors. The approach consists in casting an ejector as a pipeline network with two inlets and one outlet and allows to simulate transients characterized by pressure waves and flow reversal. The formulation combines the integral conservation of mass, momentum and energy along the primary, secondary and mixing sections with a junction model inspired by the model of 
Lang and Mindt \cite{lang}. The proposed model does not distinguish between the two streams in the mixing chamber as in the 1D models in \cite{BANASIAK20112235, delvalle2012}, but instead focuses on the local conservation of integral quantities (mass, momentum and energy) in space \emph{and} time as in classic unsteady models for pipeline networks \cite{reigstad2014, morin2015, lang}.


We validate the model in terms of entrainment ratios in steady conditions on experimental data by Besagni \emph{et al} \cite{BESAGNI2015697} and Mazzelli \emph{et al} \cite{MAZZELLI2015305}. We then use the proposed 1D approach to analyze the choking mechanism predicted by the integral formulation. Finally, we benchmark our model against 2D RANS simulations in steady and transient conditions.

The model derivation is presented in Section \ref{sec:2}. Section \ref{sec:3} reports on the numerical methods for the implementation of the 1D model while Section \ref{sec:calibration} describes the model calibration methods based on either local or global quantities. Section \ref{sec:CFD_settings} reviews the details of the CFD simulation. Section \ref{sec:5} collects the results of the experimental and numerical validation. Section \ref{sec:6} closes the paper with conclusions and future work.

\section{Model definition}\label{sec:2}

\begin{figure}[!htb]
\center
\includegraphics[width=0.8\columnwidth]{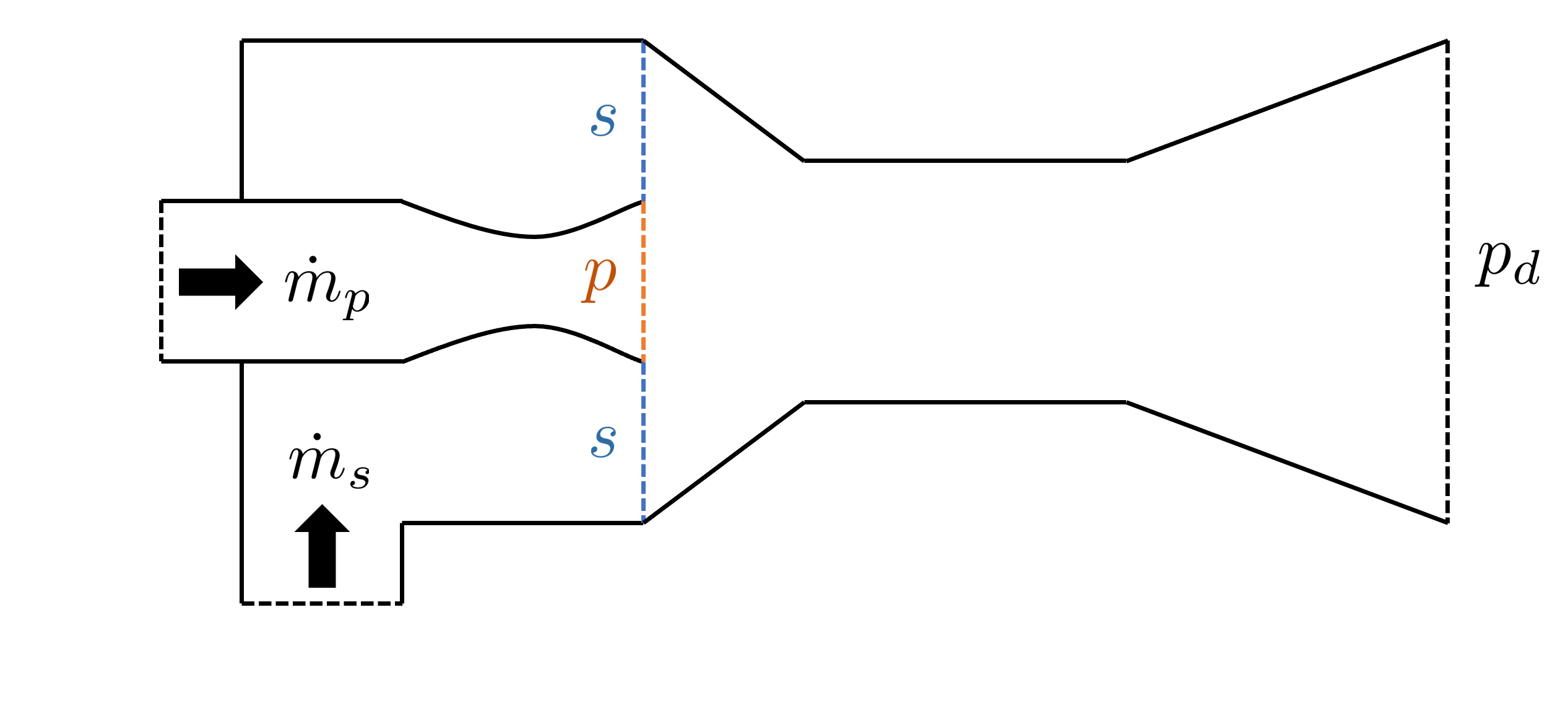}
\caption{Schematics of a supersonic ejector. A high pressure flow is expanded through a primary nozzle, entraining a secondary flow into the mixing duct.}
\label{fig:ejector_schematic}
\end{figure}

The main parameters and the key sections of an ejector are recalled in the schematic of Figure \ref{fig:ejector_schematic}. A primary flow with total pressure $p_{t,p}$ and total temperature $T_{t,p}$ expands through a nozzle with exit cross-section area $A_p$, delivering a mass flow rate $\dot{m}_p$. This entrains a secondary flow, with mass flow rate $\dot{m}_s$, total pressure $p_{t,s}$ and total temperature $T_{t,s}$, through a section of area $A_s$. The two streams mix throughout a mixing section and discharge at the outlet, at static pressure $p_d$, after passing through a diffuser.  

\begin{figure}[!htb]
\center
\includegraphics[width=\columnwidth]{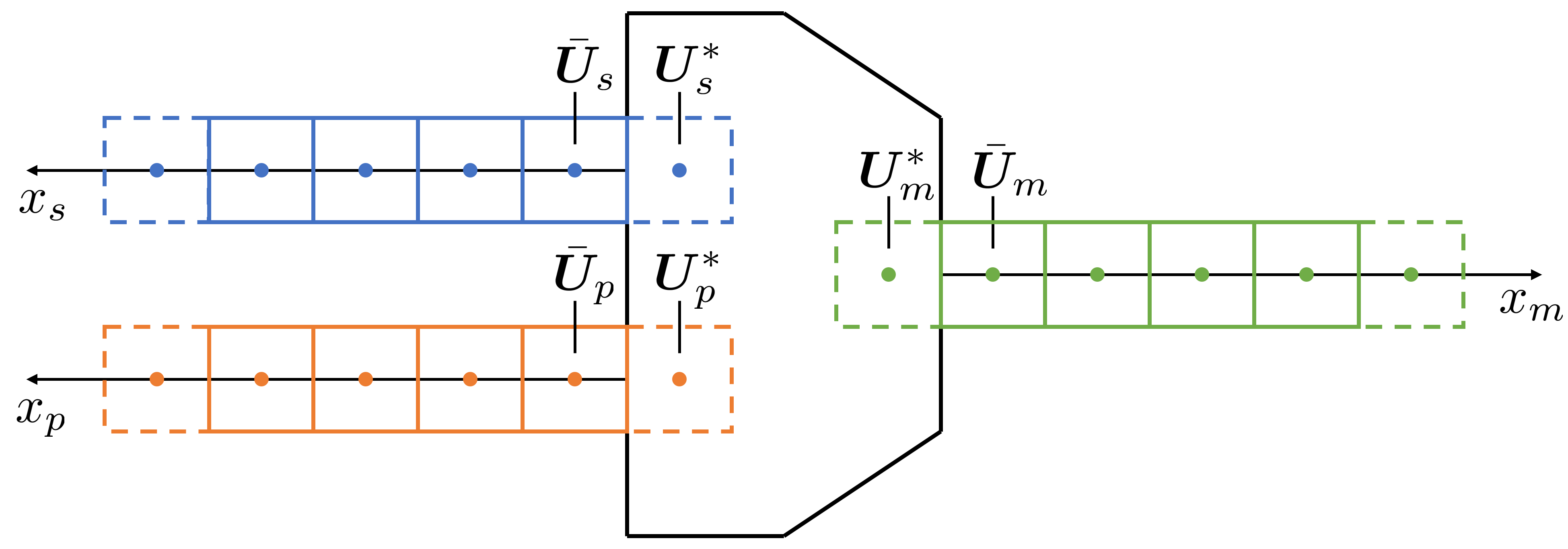}
\caption{Definitions of `initial' and `star' states in the 1D ejector model pictured as three 1D domains joined through a junction model.}
\label{fig:junction_schematic}
\end{figure}

In what follows, we assume that the operating fluid is an ideal gas and the flow is adiabatic. We restrict our model to conditions at which the primary flow is choked and the two streams are parallel when entering the mixing region. We then picture the ejector as a pipeline network consisting of two inlets (for the primary and the secondary flows) and one outlet. Each of these pipes is modeled as a 1D domain, as shown in Figure \ref{fig:junction_schematic}.

Naturally, the physics of the mixing process in a supersonic ejector differ significantly from that occurring in the junction of a pipe network. Nevertheless, considering that we focus on integral quantities and not on the detailed flow field, this is not a limit. On the contrary, the results in this work show that the approximation is appropriate.

We denote the stream-wise coordinate as $x$ in the three domains. Within each of these, we describe the flow using the quasi-1D Euler equations with source terms accounting for the cross-sectional area variation $A(x)$ \cite{toro} and friction losses:

$$
\partial_t
\begin{pmatrix}
\rho \\ \rho u \\ \rho e_t
\end{pmatrix}
+ \partial_x
\begin{pmatrix}
\rho u \\ \rho u^2 + p \\ \rho u h_t
\end{pmatrix}
= - \frac{\partial_x A}{A}
\begin{pmatrix}
\rho u \\ \rho u^2 \\ \rho u h_t
\end{pmatrix}
- \frac{1}{2} f \frac{P}{A}
\begin{pmatrix}
0 \\ \rho \left|u\right|u \\ 0
\end{pmatrix}
$$ where $\rho$ is the density, $u$ the (average) velocity, $e_t = C_V T + 0.5 u^2$ the total specific internal energy, $p$ the pressure, $h_t = e_t + p/\rho$ the total specific enthalpy, $f$ a friction factor, $P$ is the cross-section's perimeter and $A$ is the cross-section's area. The reader is referred to the list of symbols for the nomenclature.

These equations are typically used in nozzle flows \cite{anderson}. All variables depend on the stream-wise coordinate $x$, except for the friction factor $f$ which is computed with a correlation inspired by Blasius' formula \cite{bird2002transport}:

\begin{equation}
f = c \left(\textit{Re}_d\right)^{-0.25}\,,
\end{equation} with the Reynolds number $\textit{Re}_d$ based on the local hydraulic diameter. The value for $c$ is $0.0791$ for turbulent flow in smooth pipes, but in this work this coefficient is considered as a model parameter that needs to be calibrated (see Section \ref{sec:calibration}). We allow this coefficient to differ between the primary, secondary and mixing sections and use the subscripts $p,s,m$ respectively (thus we have $c_p$, $c_s$ and $c_m$). This allows for tuning the losses in each line separately, hence accounting for different local losses such as, e.g., bends or valves. Although the current model can not localize these losses, this term allows to distribute their impact over the whole domain.

The quasi 1-D equation can be more concisely written in terms of state vector $\bm{U}=[\rho, \rho u, \rho e_t]^T$ and appropriate flux function $\bm{F}(\bm{U})$ and source function $\bm{S}(\bm{U})$:

\begin{equation} \label{eq:generalform}
\partial_t \bm{U} + \partial_x \bm{F} \left(\bm{U}\right)= \bm{S}\left(\bm{U}\right)\,.
\end{equation}

All the relevant variables can be computed from the state vector (see \ref{AppA}).

\begin{figure}[!htb]
\center
\begin{subfigure}[t]{\linewidth}
	\includegraphics[width=0.8\columnwidth]{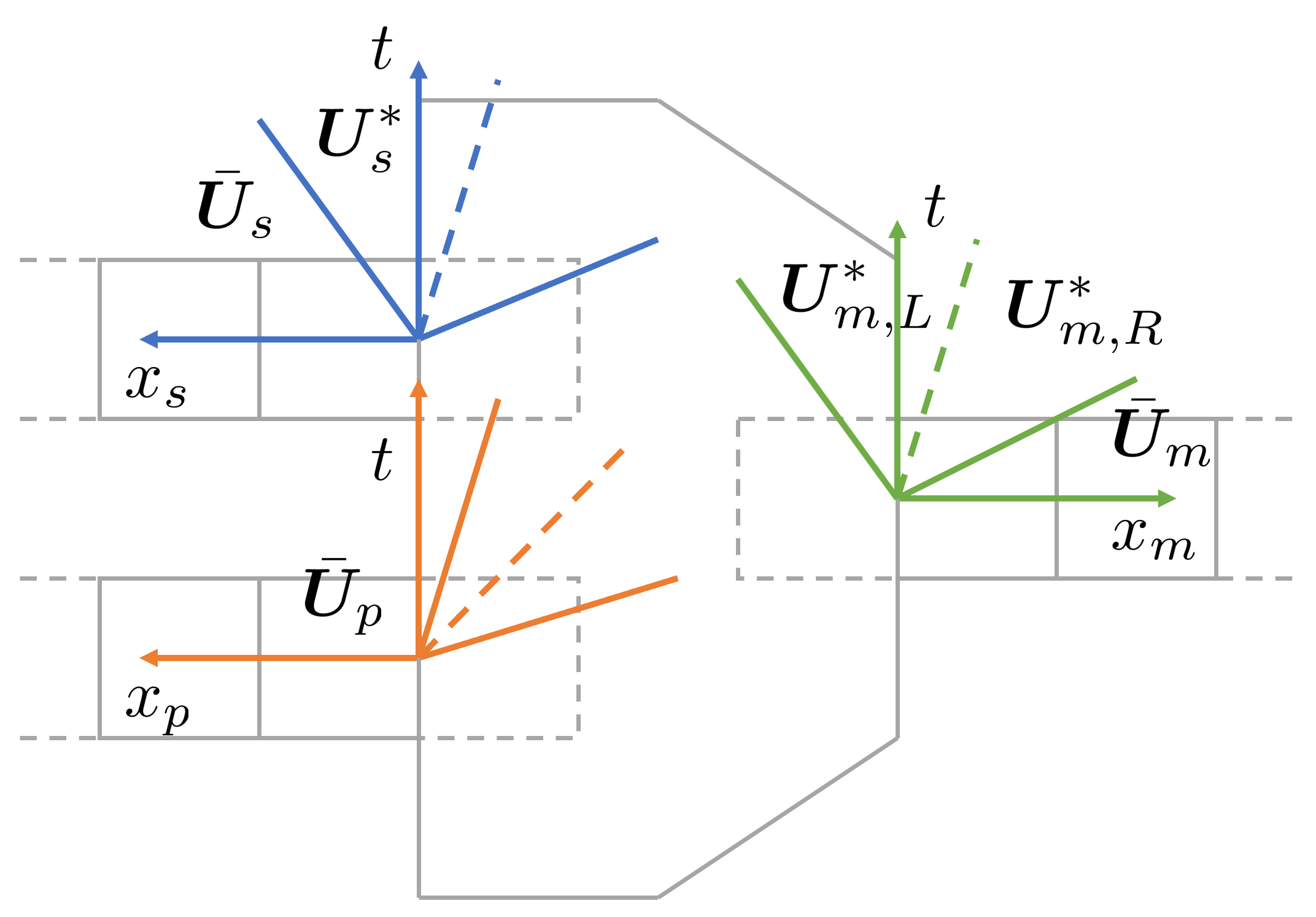}
	\subcaption{Normal operation}
	\label{fig:characteristics_normal}
\end{subfigure}
\\
\begin{subfigure}[t]{\linewidth}
	\includegraphics[width=0.8\columnwidth]{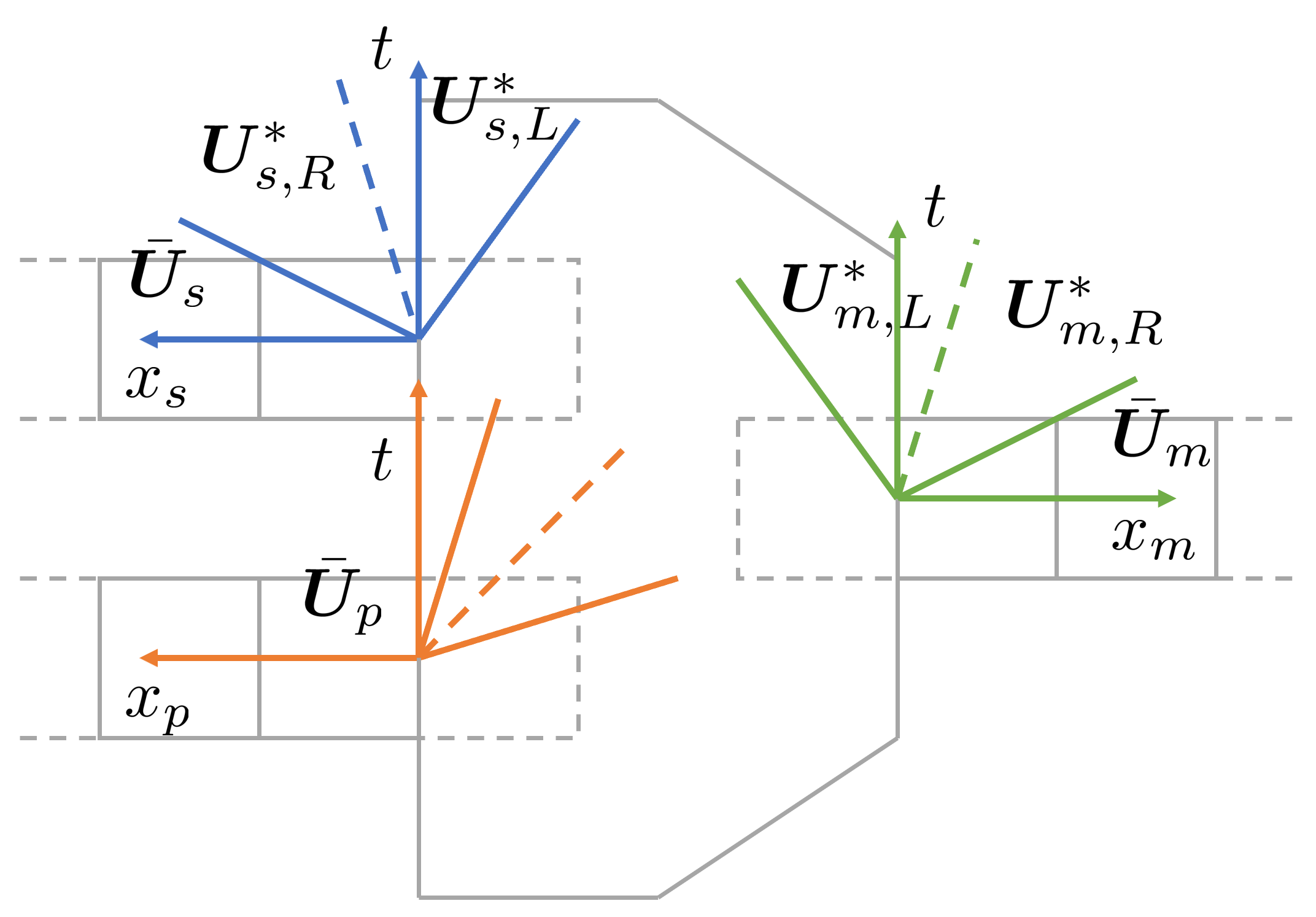}
	\subcaption{Secondary backflow}
	\label{fig:characteristics_backflow}
\end{subfigure}
\caption{Schematic of wave structures at the junction in normal (top) and backflow (bottom) operation with a supersonic primary flow and subsonic secondary and mixed flow. Solid lines represent pressure waves, dashed lines represent contact lines. The secondary and mixing boundary lie inside a star region.}
\label{fig:characteristics}
\end{figure}

The solution of \eqref{eq:generalform} is coupled with a mixing model inspired by the junction model by Lang and Mindt \cite{lang}. This treats the junction as a generalized Riemann problem, which reduces to the classic shock tube problem \cite{toro} if two pipes are considered. The extension to an arbitrary number of pipes is carried out by imposing a pressure wave in each pipe and a contact discontinuity in the outlet pipes. Figure \ref{fig:characteristics_normal} illustrates the wave structure in the normal operating conditions of an ejector: the characteristic lines of the primary supersonic flow are directed downstream; the subsonic secondary and mixed flow admit one characteristic line directed upstream. Figure \ref{fig:characteristics_backflow} shows the same schematic in the case of backflow in the secondary inlet.

The definition of the (time dependent) boundary conditions for the three domains is computed by solving a Riemann problem for each pipe. These are characterized by `star' states $\bm{U}^*_i$ linked to the initial states $\bar{\bm{U}_i}$ by the Rankine-Hugoniot relations \cite{toro}. In both cases, we have $i = p,s,m$.

Following \cite{lang}, the jump relations are

\begin{align} \label{eq:rankine_compression_velocity}
    u^*\left(p^*, \bar{\bm{U}}\right) = & \bar{u} - \left(p^*-\bar{p}\right)\sqrt{\frac{1-\mu^2}{\bar{\rho}\left(p^*+\mu^2 \bar{p}\right)}}\\\label{eq:rankine_compression_density}
    \bar{\rho}^*\left(p^*, \bar{\bm{U}}\right) = & \bar{\rho} \left(\frac{p^* + \mu^2 \bar{p}}{\mu^2 p^* + \bar{p}}\right)
\end{align} in case of a compression, i.e. $p^*>\bar{p}$, and 

\begin{align} \label{eq:rankine_expansion_velocity}
    u^*\left(p^*, \bar{\bm{U}}\right) = & \bar{u} - \frac{2 \bar{a}}{\gamma-1} \left(\left(\frac{p^*}{\bar{p}}\right)^\frac{\gamma-1}{2\gamma}-1\right)\\
    \label{eq:rankine_expansion_density}
    \bar{\rho}^*\left(p^*, \bar{\bm{U}}\right) = & \bar{\rho} \left(\frac{p^*}{\bar{p}}\right)^\frac{1}{\gamma}
\end{align} in case of an expansion, i.e. $p^*<\bar{p}$. In these relations, $\mu = (\gamma-1)/(\gamma+1)$. A contact discontinuity splits the star region of the outlet pipes (cf. figure \ref{fig:characteristics}) into left ($\bm{U}_L^*$) and right ($\bm{U}_R^*$) states. These have equal velocity and pressure but differ in the density by an (unknown) jump $\Delta \rho$, i.e. $\rho_L^* = \rho_R^* + \Delta\rho$.

These conditions must be combined with coupling conditions ensuring conservation of mass, momentum and energy. To simplify the problem, the formulation in \cite{lang} assumes equal total enthalpy for all inlets and outlets. In this work, we release this assumption and consider the complete balance equations. In normal operating conditions, these read

\begin{align} 
    0 &= \dot{m}_{p} + \dot{m}_{s} + \dot{m}_{m} \label{eq:coupl_mass} \\
    0 &= \eta_m \left(\dot{m}_{p} u_p + A_p p_p+ \dot{m}_{s} u_s + A_s p_s\right) - \dot{m}_{m} u_m - A_m p_m \label{eq:coupl_momentum} \\
    0 &= \dot{m}_{p} h_{t,p} + \dot{m}_{s} h_{t,s} + \dot{m}_{m} h_{t,m} \label{eq:coupl_energy}
\end{align} where all variables refer to `star' conditions but the stars are omitted for conciseness. In the momentum balance, the mixing efficiency $\eta_m$ has been introduced to account for momentum losses due to, e.g., force exchanges with solid walls. In case of backflow, the energy equation is simplified as $h_{t,p}=h_{t,s}=h_{t,m}$ since the junction is supplied by one single line.

The lack of an unsteady term in the conservation of mass implies that the mixing is inertia-less and no mass accumulation occurs in the junction. Moreover, variables in the outlet pipe are to be interpreted as `averaged' quantities, not implying the complete mixing between the two streams. Finally, the momentum equation focuses on the stream-wise coordinate because of the parallel streams assumption while the total enthalpy conservation stems from the adiabatic assumption.

In the numerical implementation of the model, every time step requires the solution of equations \eqref{eq:rankine_compression_velocity}-\eqref{eq:coupl_energy} for the unknown $p_s^*$, $p_m^*$ and $\Delta \rho_m$. An additional jump in density $\Delta \rho_s$ for the secondary line appears in the case of backflow, bringing the total number of unknowns to 4. In these conditions, we simplify the energy conservation equation by imposing more restrictive equality of total enthalpy as in \cite{lang}. The resulting star states $\bm{U}*$ are then provided as boundary conditions to advance the numerical integration of \eqref{eq:generalform}. The next section introduces the details for the numerical implementation.

\section{Numerical methods}\label{sec:3}
\subsection{0D initialisation} \label{sec:steady_solver}

The 1D model described in the previous section can be easily reduced to a 0D model in steady and frictionless conditions. In this case, the flow field is fully determined by the coupling equations and the boundary conditions. The 1D steady Euler equations in each pipe ensure the conservation of total pressure and total enthalpy. Moreover, in steady conditions, the Mach number can be computed in the whole domain using the classic area-Mach relation \cite{anderson} if it is known in only one section:

\begin{equation} \label{eq:areaMach}
    \left(\frac{A_1}{A_2}\right)^2 = \frac{M_2^2}{M_1^2}\left[\frac{\left(1+\frac{\gamma-1}{2}M_1^2\right)}{\left(1+\frac{\gamma-1}{2}M_2^2\right)}\right]^\frac{\gamma+1}{\gamma-1}
\end{equation} where the indices $1,2$ indicate two arbitrary cross-sections between which the flow is isentropic.

The resulting 0D model allows for initializing the 1D (unsteady) computations with a solution that respects the junction conditions, hence avoiding numerical instabilities and speeding up convergence to steady state conditions.

\begin{figure}[!htb]
\center
\includegraphics[width=0.7\columnwidth]{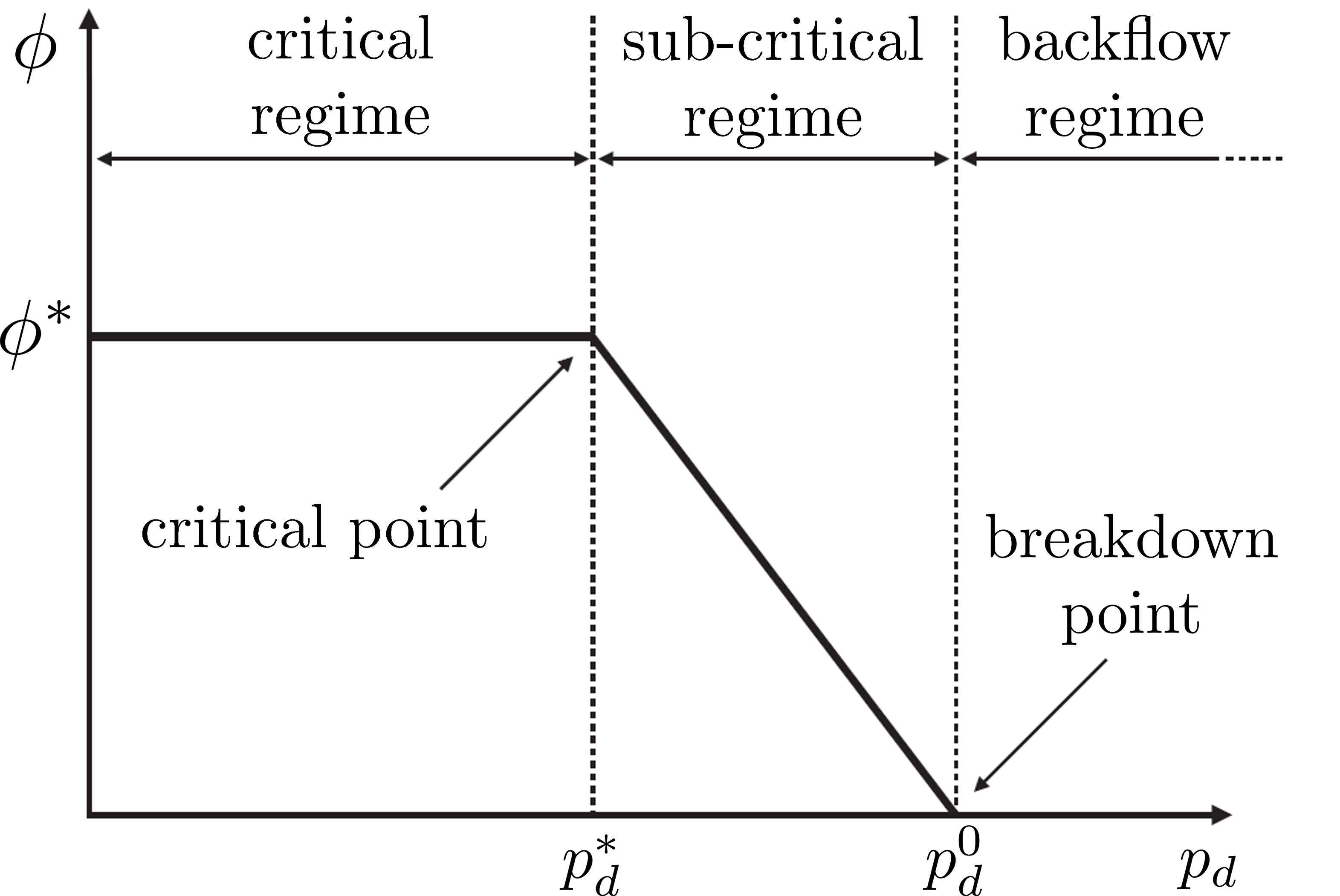}
\caption{A typical ejector operating curve. The entrainment ratio $\phi$ is independent of the back pressure $p_d$ in the critical regime, where the mixed flow is choked. Higher back pressures decrease the secondary flow rate and eventually leads to backflow.}
\label{fig:operating_curve_schematic}
\end{figure}

All variables related to the primary flow can be easily computed from the choked throat using equation \eqref{eq:areaMach} and the inlet conditions. The computation in the other domains depends on the operating conditions of the ejector. These are recalled in figure \ref{fig:operating_curve_schematic} in terms of entrainment ratio $\phi=\dot{m}_s/\dot{m}_p$ versus back pressure $p_d$ as customarily done in the literature of ejectors. 

For a given ejector ($\bm{A} = \left[A_p,A_s,A_m\right]^T$) and operating conditions ($\bm{c} = \left[p_{t,p}, T_{t,p}, p_{t,s}, T_{t,s}, p_d\right]^T$), the identification of the operating regime can be carried out given the critical point (at back pressure $p^*_d$) and the breakdown point (at back pressure $p^0_d$). Therefore, these conditions are calculated first.  

The critical operating point is characterized by the maximum pressure for which the mixing pipe is choked. As illustrated in Section \ref{sec:choking_mech}, this choking is to be interpreted in the compound choking formalism (see \cite{LAMBERTS2018_compound}). Therefore, the Mach number $M_m(x)$ can be calculated in the same way as for the primary nozzle. In the critical point, the flow reaches $M_m(x)=1$ in the mixing pipe and is subsonic in the diffuser. The associated secondary Mach number $M_s$ can be found with an iterative procedure, as the solution of an equation of the kind $\mathcal{R}(M_s)=0$.
For given $M_s$ and stagnation states at the secondary inlet, one can compute all flow variables (cf. \ref{AppA}) related to the secondary flow at the junction. Conservation of mass \eqref{eq:coupl_mass} and energy \eqref{eq:coupl_energy} then provide $\dot{m}_m$ and $h_{t,m}$ since both inlet states are now known. Then, these variables can be introduced in the momentum equation \eqref{eq:coupl_momentum}: the function $\mathcal{R}(M_s)$ is the residual in this equation. It is worth noting that the mixing efficiency $\eta_m$ influences the solution of the 0D model.

The breakdown point is characterized by no secondary flow, so the secondary pipe has zero velocity and uniform total pressure and total temperature. The mass flow rate and total enthalpy are identical in the primary and mixing pipes, leaving the Mach number $M_m$ as the only unknown. This is found with a similar iterative procedure as for the critical point, leading to the breakdown pressure $p_d^0$.

Given $p^*_d$ and $p^0_d$, the computation of the remaining variables depends on the operating regime and is described in what follows.

\paragraph*{Critical regime: $p_d < p_d^*$} \hfill

In the critical regime, the flow is choked but the static pressure is lower than the one in isentropic conditions. Consequently, the flow is supersonic in the diffuser and expands further: shocks must occur in the diffuser to let the pressure reach the prescribed back pressure. These appear as a sequence of oblique shocks in a real ejector\cite{BARTOSIEWICZ200556}, but the proposed 1D formulation pictures them as a single normal shock. This is a classic assumption in the LPM of ejectors \cite{eames1995,chen2013,METSUE2021121856}.

In these conditions, all variables of interest depend on the position of this fictitious normal shock. We herein denote as $x_{sh}$ its distance from the diffuser inlet. For a given $x_{sh}$, the geometry of the diffuser prescribes the cross-sectional area $A_{sh}$ at which this occurs, while the area-Mach equation \eqref{eq:areaMach} prescribes the Mach number $M_{up}$ upstream of the shock. The normal shock equations then provide the Mach number $M_{down}$ and pressure $p_{down}$ downstream of the shock:

\begin{align}
    M_{down}^2 &= \frac{1+\frac{\gamma-1}{2} M_{up}^2}{\gamma M_{up}^2-\frac{\gamma-1}{2}}\\
    \frac{p_{down}}{p_{up}} &= 1 + \frac{2\gamma}{\gamma+1}\left(M_{up}^2-1\right)\,.
\end{align}

Downstream the shock, the subsonic flow decelerates isentropically to a Mach number $M_d$ at the diffuser exit (cf. equation \ref{eq:areaMach}). The static pressure $p_d$ follows from the conservation of total pressure after the shock. One can thus iterate on $x_{sh}$ until the resulting back pressure is equal to the imposed one. 

\paragraph*{Sub-critical regime: $p_d^* < p_d < p_d^0$} \hfill

In the sub-critical regime the flow is subsonic in both the secondary and the mixing pipes. Therefore, on must iterate on both Mach numbers $M_s$ and $M_m$. These are identified by ensuring the compliance with the momentum equation \emph{and} the matching with the back pressure. 

\paragraph*{Backflow regime: $p_d^0 < p_d$} \hfill

In this work, we restrict the modeling to the backflow in the secondary pipe (cf. the characteristic lines in figure \ref{fig:characteristics}). In principle, the flow can reverse in any section at sufficiently high back pressure, but this might conflict with the assumption of choked primary flow. The (unknown) Mach numbers $M_s$ and $M_m$ are below unity, so they are iterated upon as in the normal sub-critical regime. The reversed secondary flow has two consequences on the calculation procedure: (1) the energy conservation equation in the junction changes to equality of total enthalpy in all pipes (cf. section \ref{sec:2}) and (2) the stagnation pressure $p_{t,s}$ at the inlet is now the static discharge pressure of the reversed secondary flow. The mass flow rate $\dot{m}_s$ can be calculated from the guessed Mach number $M_s$, the total temperature $T_{t,s} = T_{t,p}$ and the static discharge pressure $p_{t,s}$ with gas dynamic relations (cf. \ref{AppA}). The remaining calculation is similar to the sub-critical regime: the conservation of mass provides $\dot{m}_m$, which allows calculating the residual on the momentum equation, and the static back pressure from the guessed Mach number $M_m$ and the known total temperature $T_{t,m}=T_{t,p}$.

\subsection{Unsteady computations}

The system of equations in \eqref{eq:generalform} is solved using a classic finite volume approach. The source term is treated explicitly and the time marching is carried out with a first order Euler scheme (cf. chapter 15 of \cite{toro}):

\begin{equation} \label{eq:num_scheme}
    \bm{U}_i^{n+1} = \bm{U}_i^n - \frac{\Delta t}{\Delta x} \left(\bm{F}_{i+\frac{1}{2}}^n - \bm{F}_{i-\frac{1}{2}}^n\right) + \Delta t \bm{S}_i^n\,,
\end{equation}

where the subscript $i$ denotes the cell index in space and $n$ denotes the index in time. The fluxes are calculated with a first order Roe scheme:

\begin{equation} \label{eq:Roe_flux}
    \bm{F}_{i+\frac{1}{2}} = \frac{1}{2} \left(\bm{F}_i + \bm{F}_{i+1}\right) - \frac{1}{2} \sum_{k=1}^3\tilde{\alpha}_k \left| \tilde{\lambda_k} \right| \tilde{\bm{\text{K}}}_k\,,
\end{equation} where $\tilde{\alpha}_k$ denotes the wave strength and $\tilde{\lambda_k}$ and $\tilde{\bm{\text{K}}}_k$ denote the eigenvalues and eigenvectors of the system. These are recalled in \ref{AppB} for completeness, but the interested reader is referred to chapter 11 of \cite{toro} for more details. Boundary conditions are handled with ghost cells as proposed in chapter 7 of \cite{lev02} (cf. figure \ref{fig:junction_schematic}). The boundary fluxes can thus be calculated with the internal scheme, which requires the states at both sides of the interface.

The junction model only provides boundary conditions and does not impose any requirements on the numerical scheme. The first order Roe scheme was chosen in this work for its low diffusion and its relative simplicity.

\section{Model calibration} \label{sec:calibration}

The current model can be tuned with 4 coefficients (cf. section \ref{sec:2}) which we group into a single vector for conciseness: $\bm{w} = \left[\eta_m, c_p, c_s, c_m \right]^T$. These coefficients must be adjusted to match the model prediction with experimental or numerical data. Depending on the available data, the calibration might be based on global quantities (e.g. the entrainment ratio) or locally averaged quantities (e.g. the downstream evolution of the momentum).

To the authors' knowledge, no experimental data is available in the literature concerning locally averaged quantities. Most of the available experimental works provide data on entrainment ratio $\phi$ as a function of various operating conditions, while only a few authors give both mass flow rates. On the other hand, 2D or 3D simulations can easily provide data for the local validation of the proposed model. 

Regardless of the approach, the calibration procedure is a regression problem in which a cost function $J(\bm{w})$ must be minimized. In a global approach, this cost function might take the form 

\begin{equation}
J\left(\bm{w}\right) = 
\begin{cases}
\sum_{j=1}^{n_p} \left(\dot{m}_{p,j}-\dot{m}_{p,j}^d\right)^2 + \left(\dot{m}_{s,j}-\dot{m}_{s,j}^d\right)^2 & \text{ if available}\\
\sum_{j=1}^{n_p} \left(\phi_j-\phi_j^d\right)^2 & \text{ otherwise}
\end{cases}
\label{eq:calibration_global}
\end{equation} where $n_p$ is the number of available data points, the superscript $d$ refers to available data and the j-th prediction is provided by the model at all the conditions at which the data was measured:
\begin{equation}
    \dot{m}_{p,j}, \dot{m}_{s,j} = f\left(\bm{A}_j, \bm{c}_j; \bm{w}\right)\,.
\end{equation}

Likewise, in a local approach the cost function could be

\begin{align}
J\left(\bm{w}\right) = \sum_{i=p,s,m} \sum_{n=1}^{n_t} \nonumber
\Bigg(&\alpha_1\left|\left|\dot{\bm{m}}_{i,n}-\dot{\bm{m}}_{i,n}^d\right|\right|_2^2 + \alpha_2\left|\left|\bm{\mathcal{M}}_{i,n}-\bm{\mathcal{M}}_{i,n}^d\right|\right|_2^2 \\
& + \alpha_3\left|\left|(\dot{\bm{m}} \bm{h}_{t})_{i,n}-(\dot{\bm{m}} \bm{h}_{t})_{i,n}^d\right|\right|_2^2
\Bigg)
\label{eq:calibration_local}
\end{align} where $||\bullet||_2$ denotes the $l_2$ norm of a vector, the coefficients $\alpha_1,\alpha_2,\alpha_3$ can be used to define different weights and the variables $\dot{\bm{m}}_{i,n}$, $\bm{\mathcal{M}}_{i,n}$ and $(\dot{\bm{m}} \bm{h}_{t})_{i,n}$ denote the vectors of mass, momentum and enthalpy flow rates in all cells of domain $i$ at time step $n$. These quantities are defined by integrating over the cross stream section as follows

\begin{equation}
\label{m_RATE}
    \dot{m}(x) = \int_A \rho u dA
\end{equation}
\begin{equation}
    \label{M_RATE}
    \mathcal{M}(x) = \int_A (\rho u^2 + p) dA
\end{equation}
\begin{equation}
\label{H_RATE}
    \dot{m} h_t(x) = \int_A \rho u h_t dA
\end{equation}

Of course many other cost functionals can be defined and one might not necessarily restrict to single objective optimization. Similarly, one might set up the closure problem in terms of more complex friction losses (e.g. nonuniform friction coefficients or closure parameters depending on the operating conditions) but none of these extensions fall within the scope of this work. 

In both global and local calibrations, the minimization of the cost functions was carried out using a the L-BFGS-B algorithm \cite{byrd1995limited}, as implemented in the Python package \textsc{Scipy} \cite{scipy}, with the following bounds for the closure parameters: $c_p,c_s,c_m \in [0,0.5]$ and $\eta_m\in [0.85, 1.05]$. It is worth noting that a value $\eta_m>1$ can occur if $A_m>A_p+A_s$ due to, e.g., the finite thickness of the surfaces separating the primary from the secondary flows.


\begin{figure*}[!htb]
    \centering
    \includegraphics[width=\linewidth]{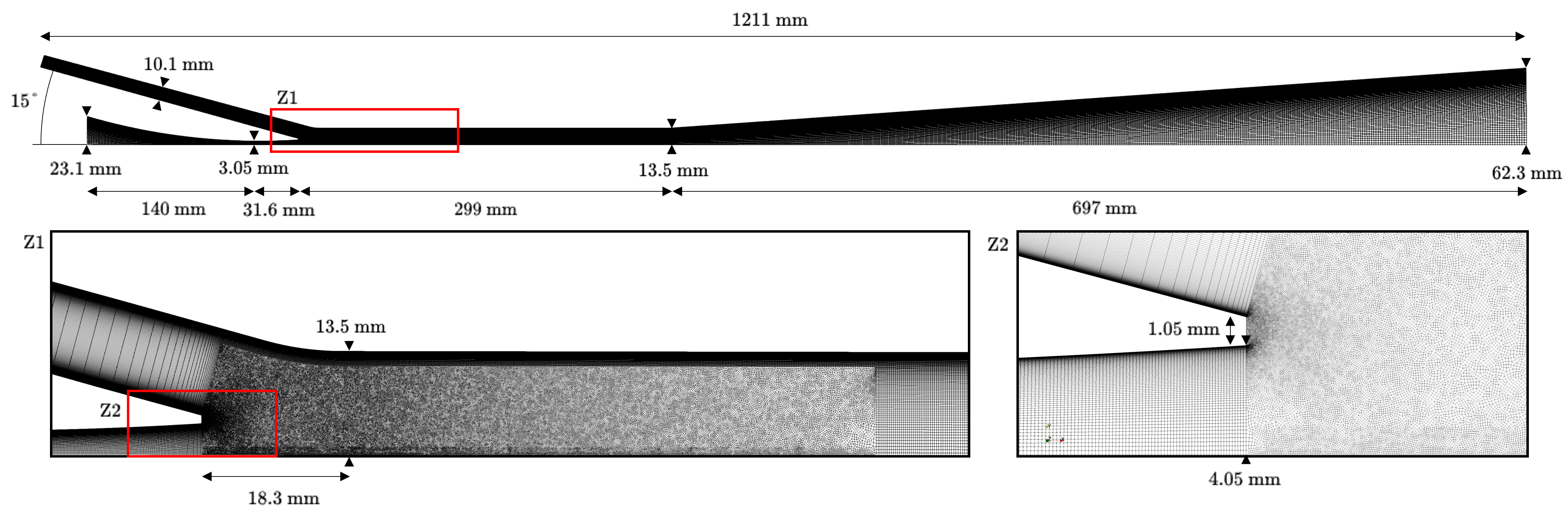}
    \caption{Geometry and mesh used in the 2D CFD simulation. The geometry is planar and has a width of $48.8$ mm. The 2D mesh is mostly structured and contains $338164$ cells. A symmetry boundary condition is imposed on the bottom plane.}
    \label{fig:CFD_mesh}
\end{figure*}

\section{2D CFD Benchmark} \label{sec:CFD_settings}

2D URANS computations were performed in both steady and unsteady conditions, to benchmark the proposed 1D model. The numerical settings, herein described, are the same in the two cases. The ejector geometry and the computational domain is adapted from Lamberts et al. \cite{LAMBERTS201723} and is shown in figure \ref{fig:CFD_mesh}. The mesh is mostly structured, except near the exit of the primary nozzle. The $k \text{-} \omega$ SST turbulence model is used as it is known to be well suited for ejectors \cite{BARTOSIEWICZ200556}. The transient, compressible solver \textit{rhoCentralFoam} from \textit{OpenFOAM v9} \footnote{\url{https://openfoam.org/version/9/}} has been used for both simulations.

Table \ref{tab:BC_CFD} collects the kind of applied boundary conditions. The total pressure, the total temperature and the turbulence intensity \cite{openfoam_TI_BC} are imposed at the inlets, with zero gradient for the velocity. The mass flow rates are thus a result from the calculation. The detailed operating conditions are reported in the next section for the different simulations. For all simulations, the mixing length for computing the turbulence dissipation $\omega$ and the turbulence intensity are the same as in Lamberts \cite{LAMBERTS201723}, that is $l_{mix}=0.0044$ m and $TI=5$ \%. The reader is referred to the \textit{OpenFOAM}'s user guide \cite{openfoam_lmix_BC} for the computation of the relevant turbulence quantities from these boundary conditions.

Zero gradients are also imposed at the outlet except for the pressure. The wave transmissive condition is used here to prevent excessive reflections at the inlet/outlet boundaries. These were set with a relaxation length scale of $10^{-4}$ m \cite{openfoam_wavetransmissive_BC}. The walls have no slip, are adiabatic, and use low-Reynolds wall functions \cite{openfoam_k_wall_BC, openfoam_omega_wall_BC}. The mesh is resolving the boundary layer with $y^+< 1$.

\begin{table}[!htb]
    \centering
    \begin{tabular}{|c|cccc|}
        \hline
         & primary & secondary & outlet & walls \\
         & inlet & inlet & & \\
        \hline
        $p$ & $p_{t,p}$ & $p_{t,s}$ & transmissive & $\partial_n p = 0$ \\
        $T$ & $T_{t,p}$ & $T_{t,s}$ & $\partial_n T = 0$ & $\partial_n T = 0$ \\
        $\bm{v}$ & $\partial_n \bm{v} = 0$ & $\partial_n \bm{v} = 0$ & $\partial_n \bm{v} = 0$ & $0$ \\
        $k$ & $TI$ & $TI$ & $\partial_n k = 0$ & wall function \\
        $\omega$ & $l_{mix}$& $l_{mix}$ & $\partial_n \omega = 0$ & wall function \\
        \hline
    \end{tabular}
    \caption{Boundary conditions used for the 2D CFD. The values for the pressure and temperature follow from the operating conditions.}
    \label{tab:BC_CFD}
\end{table}

\section{Results}\label{sec:5}
The model is first validated globally against experimental data from Mazzelli \cite{MAZZELLI2015305} and Besagni \cite{BESAGNI2015697} in Section \ref{6p1}. The results from this 0D validation are further analyzed to reveal the choking mechanism according to the proposed model in section \ref{sec:choking_mech}. Section \ref{sec6p3} and \ref{6p4} reports on the benchmarking in steady and transient conditions.

\subsection{Global validation}\label{6p1}

We begin with the model calibration and testing using the experimental data by Mazzelli et al. \cite{MAZZELLI2015305} concerning a planar and supersonic ejector. The dimensions of the key sections are reported in Table \ref{tab:area_mazzelli}; the reader is referred to the original publication for more details. 

\begin{table}[!htb]
    \centering
    \begin{tabular}{|c|ccc|}
        \hline
        domain & inlet & outlet & length\\
        \hline
        primary & $110.25 $ & $8.00 $ & 422\\
        secondary & $2 \times 62.14 $ & $2 \times 12.72 $ & 430\\
        mixing & $27.06 $ & $130.14 $ & 1019 \\
        \hline
    \end{tabular}
    \caption{Height and length of the inlet and outlet sections for the experiments reported by Mazzelli \cite{MAZZELLI2015305}. The geometry is planar with a uniform depth of 48.8 mm. The primary throat is $5.98 \text{ mm}$ high. The secondary section is counted twice to account for both inlets. All dimensions are in mm.}
    \label{tab:area_mazzelli}
\end{table}

\begin{figure}[!htb]
    \centering
    \includegraphics[width=\columnwidth]{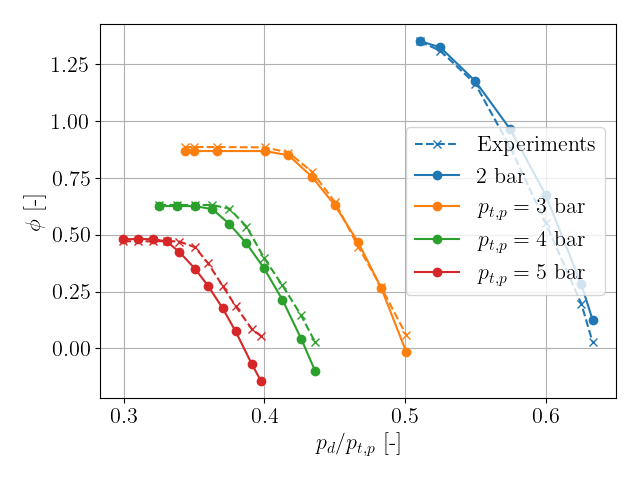}
    \caption{Comparison of experimental data (in \cite{MAZZELLI2015305}, Fig. 4) with the converged unsteady solver for different primary pressures. The different curves are obtained keeping the same pressure in the primary nozzle. Stars are used for the experimental data and circular markers for the model prediction. The model calibration yields $\eta_m = 0.85$, $c_p = 0$, $c_s = 0.50$ and $c_m = 0$ (cf. Section \ref{sec:2} and \ref{sec:calibration}).}
    \label{fig:mazzelli_exp_steady}
\end{figure}

The experiments were conducted keeping the secondary port open to the atmosphere and inlet pressure for the primary nozzle varying in the range $2$-$5$ bar. The back pressure varies to cover a primary-to-back pressure ratio $p_d/p_{t,p}$ in the range $0.3-0.7$. Figure \ref{fig:mazzelli_exp_steady} compares experimental data and model for various pressures in the primary flow. The global calibration (see Section \ref{sec:calibration}) was carried out including the entire data set (that is the four operating curves) and the resulting coefficients are $\eta_m=0.85$, $c_p=0$, $c_s=0.5$ and $c_m=0$.

The large friction factor in the secondary inlet indicates high losses in the secondary flow. This is in line with the comparatively low isentropic efficiency ($\eta_s=0.75$, versus $\eta_p=0.977$ in the primary) used by Metsue et al. \cite{METSUE2021121856} in the calibration of their model in the same data. The different losses in the two lines might be justified by the fact that the experimental set-up features a 90-degree bend downstream of the secondary pressure measurement.

After calibration, the matching with experimental data is satisfactory. Better matching could be obtained by independently repeating the calibration for each operating curve. This allows for different coefficients for each operating curve and the derivation of a link between coefficients and operating conditions. This extension of the model calibration will be addressed in a future contribution.

As a second test case, we consider the experimental data by Besagni et al. \cite{BESAGNI2015697} on a planar and sonic ejector. The relevant geometrical parameters are collected in Table \ref{tab:validation_besagni}. The comparison between model and experimental data on the four available points is summarized in Table \ref{tab:validation_besagni}. The relative errors are below 6 \% with the model coefficients set to $\eta_m = 1$, $c_p =c_s = c_m = 0$.

The higher mixing efficiency in this test case might be due to a better design of the mixing area and the different locations chosen to define the mixing section. This was taken at the start of the constant area section in the geometry by Mazzelli et al. \cite{MAZZELLI2015305} and right after the primary nozzle exit for the geometry by Besagni et al. \cite{BESAGNI2015697}. The much larger hydraulic diameters in all sections of the ejector could also justify the negligible friction losses in the geometry by Besagni et al. \cite{BESAGNI2015697}.

\begin{table}[!htb]
    \centering
    \begin{tabular}{|c|ccc|}
        \hline
        domain & inlet & outlet & length\\
        \hline
        primary & $21.7$ & $3.1$ & 75.0\\
        secondary & $44.5 $ & $44.5 $ & 75.0\\
        mixing & $47.6 $ & $47.6 $ & 304.8\\
        \hline
    \end{tabular}
    \caption{Height and length of the inlet and outlet sections for the experiments reported by Besagni \cite{BESAGNI2015697}. The geometry is planar with a width of 203.2 mm. The primary throat is convergent, hence its area corresponds to the outlet section. All dimensions are in mm.}
    \label{tab:area_besagni}
\end{table}

\begin{table}[!htb]
\begin{center}
\begin{tabular}{|cccccccc|}
\hline
$p_p$ & $p_s$ & $T_p$ & $T_s$ & $p_m$ & $\phi_{exp}$ & $\phi_{mod}$ & error \\

[bar] & [bar] & [K] & [K] & [bar] & [-] & [-] & [\%] \\
\hline
2.47 & 1.02 & 361 & 304 & 0.997 & 3.91 & 3.87 & 1.1 \\
2.47 & 1.02 & 359 & 302 & 0.981 & 4.17 & 4.11 & 1.4 \\
2.46 & 1.02 & 358 & 306 & 0.927 & 4.66 & 4.69 & 0.7 \\
2.46 & 1.01 & 367 & 304 & 0.927 & 5.01 & 4.73 & 5.5 \\
\hline
\end{tabular}
\caption{Operating conditions and experimental and numerical entrainment ratio on the sonic ejector by Besagni et al. \cite{BESAGNI2015697}. The mixing efficiency and the friction factors were calibrated as constants on all four points: $\eta_m = 1$, $c_p = 0$, $c_s = 0$ and $c_m = 0$.}
\label{tab:validation_besagni}
\end{center}
\end{table}

\subsection{A note on the choking mechanism} \label{sec:choking_mech}

\begin{figure}[!htb]
    \centering
    Primary Line
    \begin{subfigure}[t]{\linewidth}
		\includegraphics[width=\columnwidth]{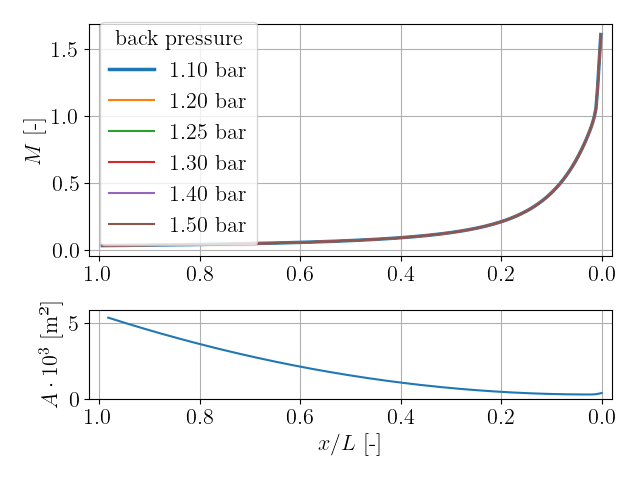}
	\end{subfigure}
	\\
	Secondary Line
	\begin{subfigure}[t]{\linewidth}
		\includegraphics[width=\columnwidth]{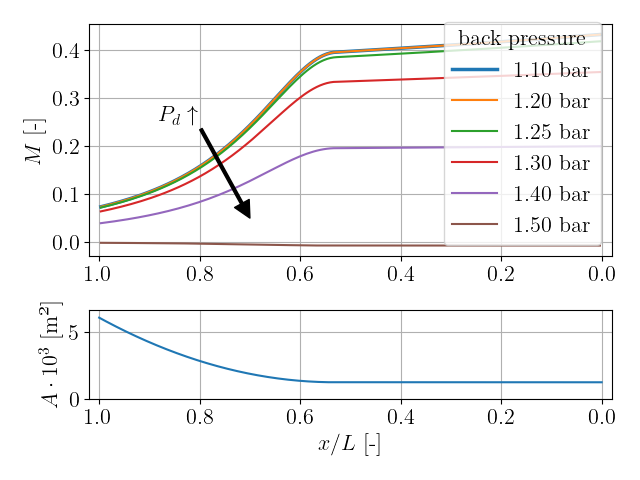}
	\end{subfigure}
	\\
	Mixing Line
	\begin{subfigure}[t]{\linewidth}
		\includegraphics[width=\columnwidth]{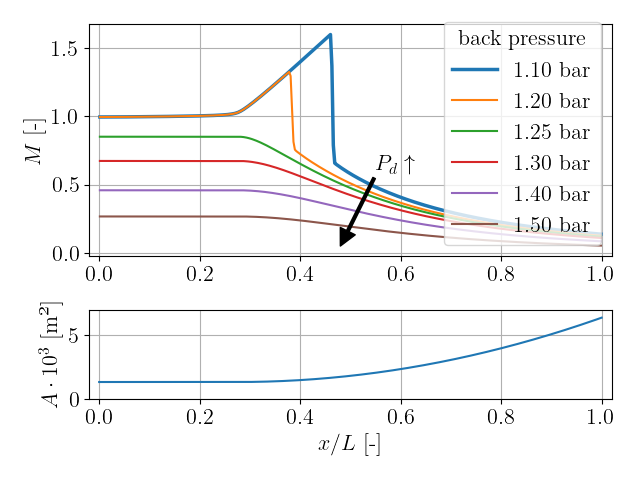}
	\end{subfigure}
	\caption{Mach number in the three 1D domains for different back pressures, as in figure \ref{fig:mazzelli_exp_steady} ($p_{t,p} = 3$ bar). The primary nozzle is choked. In the critical regime, the secondary flow is subsonic, while a shock is formed in the diffuser.}
	\label{fig:mazzelli_1D}
\end{figure}

We now consider the operating points at $p_{t,p} = 3$ bar from the dataset by Mazzelli et al. \cite{MAZZELLI2015305} (see Figure \ref{fig:mazzelli_exp_steady}) and we analyze the 1D distribution of the (local) Mach number for all the available back pressures. The results are shown in Figure \ref{fig:mazzelli_1D} for the primary, the secondary and the mixing lines, together with the corresponding area distributions. These were obtained by a quadratic fit of the originally piece-wise linear contours to avoid differentiability problems.

Since the primary flow is choked, the local Mach distribution is independent of the back pressure and reaches $M=1.61$ as expected by the isentropic condition (given that $c_p=0$).
The range of back pressures covers the entire set of regimes (see Figure \ref{fig:operating_curve_schematic}). At $p_d=1.5$ bar, the model is in the back-flow regime, and the Mach number is slightly negative in the secondary line. As visible from figure \ref{fig:mazzelli_exp_steady}, this is due to a slight miss-match between the model and the experimental data. The assumption of constant $c_s=0.5$ over the entire range of investigated conditions is too restrictive for this operating point, which is close to the break-down point (hence no secondary flow). Nevertheless, the matching in all other conditions is excellent, and Figure \ref{fig:mazzelli_1D} shows that the secondary flow is far from being choked even when the ejector reaches the critical regime and the mixed flow is choked. This is the essence of the compound choking theory: the mixture of two streams of compressible flows can choke in a channel even if one of the two is subsonic. This mechanism was recently proposed by Lamberts \emph{et al} \cite{LAMBERTS2018_compound} and explicitly embedded in a 0D model by Metsue et al. \cite{METSUE2021121856}.

Finally, the flow reaches supersonic conditions in the mixing line for the cases with back pressure of $1.1$ bar and $1.2$ bar. As also described in the model by Banasiak and Hafner \cite{BANASIAK20112235}, this results in an initial expansion in the diffuser (in this case, starting at $x/L = 0.2$), followed by a normal shock to match the back pressure. The shock moves downstream as the back pressure decreases further. It is worth stressing that this normal shock is a common modeling simplification of the actual pattern of oblique shocks encountered in a supersonic ejector \cite{BARTOSIEWICZ200556, RUSLY20051092}, and the present model naturally adjusts its location and strength.

\subsection{Steady State Benchmark versus 2D CFD}\label{sec6p3}

We compare the performances of the proposed 1D model with the 2D CFD simulations in one of the operating points in the experiments by Mazzelli \cite{MAZZELLI2015305}. The use of a 2D formulation (see section \ref{sec:CFD_settings}) for the selected geometry is questionable because it does not allow for capturing important 3D effects (e.g. friction on the side walls). This results in over prediction of the secondary flow and an erroneous computation of the critical point (see Figure 9 in \cite{MAZZELLI2015305}). Nevertheless, our interest in this benchmark was to compare the prediction of stream-wise evolution of important variables such as mass, momentum and energy fluxes with the prediction from CFD. The proposed 1D model can easily compensate for the difference with respect to the experiments by tuning the model coefficients (as proven by the global validation in Section \ref{6p1}).

We thus consider an operating condition with $p_{t,p} = 5.00$ bar, $p_{t,s} = 0.974$ bar, $p_d = 1.20$ bar and $T_{t,p}=T_{t,s}=300$ K This is taken from Table A.1 in \cite{MAZZELLI2015305} and corresponds to an on-design condition with large $p_{t,p}/p_d$ ratio. Strong shocks are thus expected in the diffuser, as confirmed by the CFD investigation in \cite{MAZZELLI2015305}. 

\begin{figure*}[!htb]
    \centering
    \includegraphics[width=\linewidth]{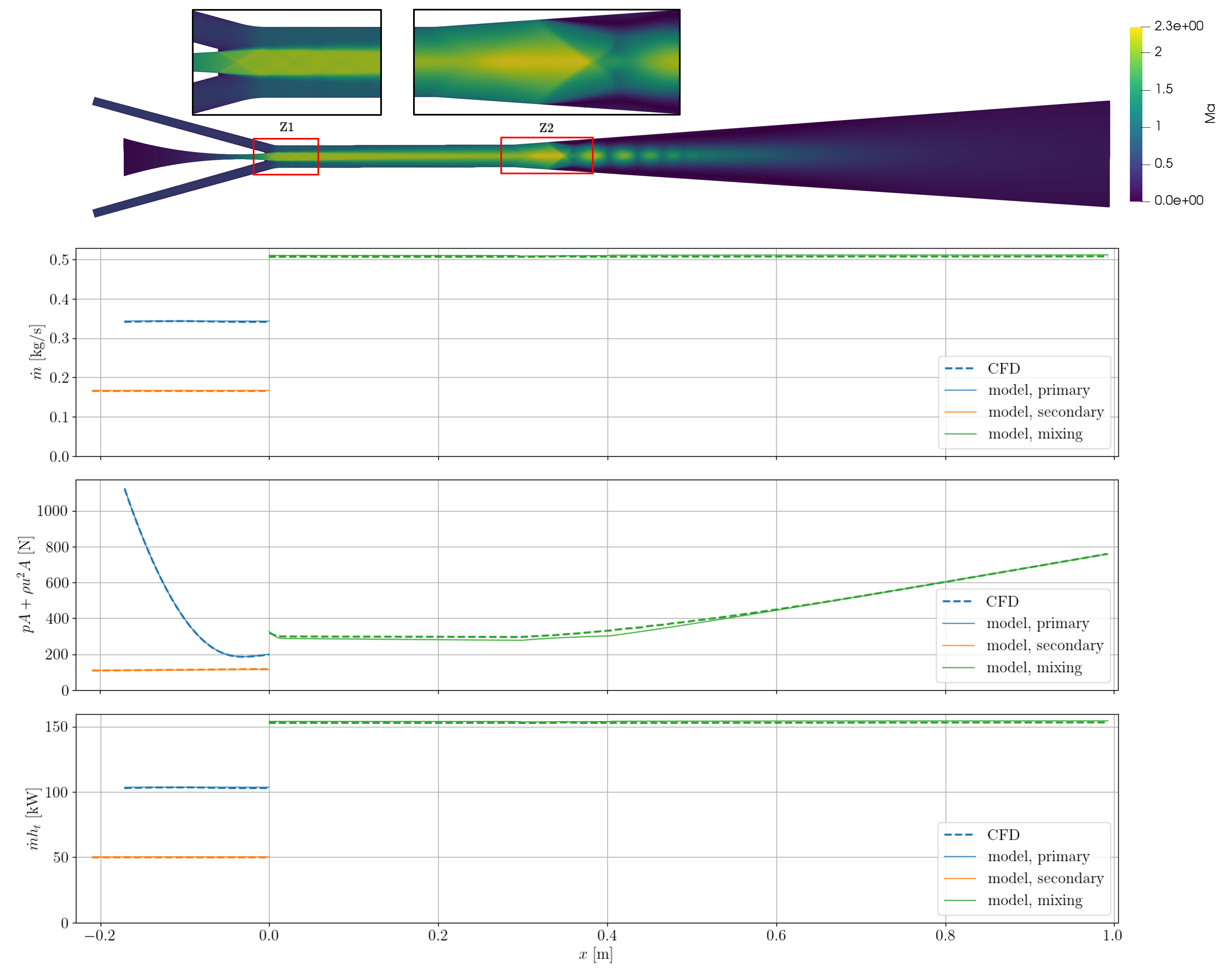}
    \caption{Local Mach number from the steady CFD simulation (top) and the comparison of the mass, momentum and energy flow rates with the 1D model (bottom). The 2D simulation reveals expansion waves in the mixing duct (Z1) and a strong shock train in the diffuser (Z2). The coupling equations \eqref{eq:coupl_mass}-\eqref{eq:coupl_energy} impose the conservation of the shown quantities at the junction ($x=0$). The computational domain has been mirrored for visualisation purposes.}
    \label{fig:CFD_Mach}
\end{figure*}

Figure \ref{fig:CFD_Mach} shows a contour of the Mach number obtained in our CFD simulation, together with a comparison of the 1D distribution of mass flow rates, momentum and total enthalpy flow rates according to the proposed 1D model. For the CFD simulations, these were computed using equations \eqref{m_RATE}-\eqref{H_RATE}.

The well-known oblique shock pattern appears at the nozzle exit (zoom region Z1) and at the diffuser's inlet (zoom region Z2). These are strong enough to trigger boundary layer separation (also reported in \cite{BARTOSIEWICZ200556} in similar conditions). The proposed model naturally places vertical shocks in these locations, and the comparison in the integral quantities is in excellent agreement. The model calibration was here carried out locally (see Section \ref{sec:calibration}) on the CFD data and yields $\eta_m = 1.04$, $c_p = 0.062$, $c_s = 0.391$ and $c_m= 0.148$. 

Finally, it is worth stressing that the integral quantities in the mixing line are to be considered as `compound' averages. The thermodynamic state of the mixture is defined regardless of whether the two streams have actually mixed (as opposed to the 1D models in \cite{BANASIAK20112235, delvalle2012}). While the integration/averaging of CFD results works well for extensive quantities such as the illustrated mass, momentum and energy rates, no similar averaging is naturally possible for intensive quantities such as pressure or Mach numbers. While the link between the local intensive quantities and the compound averages in the proposed 1D model is out of the scope of this work, the accuracy in the prediction of valuable integral quantities is proven.

\subsection{Transient Benchmark versus 2D CFD}\label{6p4}

We here consider the transient between two operating conditions for the same ejector as in the previous section. The initial operating point is the previously investigated one ($p_{t,p} = 5.00$ bar, $p_{t,s} = 0.974$ bar, $p_d = 1.20$ and $T_{t,p}=T_{t,s}=300$ K) while the final one is obtained by suddenly increasing the back pressure to $p_d=1.8$ bar while keeping everything else equal. The previous section's solution serves as the initial condition for both the CFD and 1D models.

Figure \ref{fig:CFD_p_mass} shows the static pressure contours from the 2D CFD simulation at six time steps, together with the mass flow rate predictions in both the CFD and the 1D model. The sudden increase in the back pressure triggers a strong pressure wave travelling upstream. In the 1D model, this takes the form of a normal shock while in the 2D CFD this has a curved wavefront. This is primarily due to the lower velocity near the walls. Interestingly, the velocity in the 1D model is close to the average wave velocity, and the dynamic of the back-ward travelling wave is well captured (see snapshot \ref{fig:CFD_p_mass_1.0} at time $t=1$ms). This result is remarkable if one considers the compound averaging mechanism underpinning the definition of the mixture flow quantities in the 1D model: the wave propagation speed depends on the local `global' Mach number, whose interpretation is obscure as the two streams are not completely mixed.

When the pressure wave reaches the junction, the coupling conditions require a change in the secondary flow but not in the primary one, which is assumed to be choked. Figure \ref{fig:CFD_p_mass_4.4} shows that the 1D model correctly captures the wave passage while, at the same time, the flow in the mixing pipe recovers from the sudden change and returns positive. The wave then reflects at the constant total pressure inlet and re-enters the domain as a much faster upstream travelling wave (cf. Figure \ref{fig:CFD_p_mass_7.0} at time $7$ms). This results in several oscillations before the ejector settles into the new steady state (cf. Figure \ref{fig:CFD_p_mass_7.0} at time $32.4$ms).

This test case proves that the proposed model can handle transitions between operating regimes, correctly describing traveling waves and the characteristic response time of the ejectors. Furthermore, the numerical implementation and the junction model proved sufficiently robust to handle sudden variations in the flow variables and flow reversal.


\begin{figure*}[!htb]
    \centering
    \begin{subfigure}[t]{\linewidth}
		\includegraphics[width=\columnwidth]{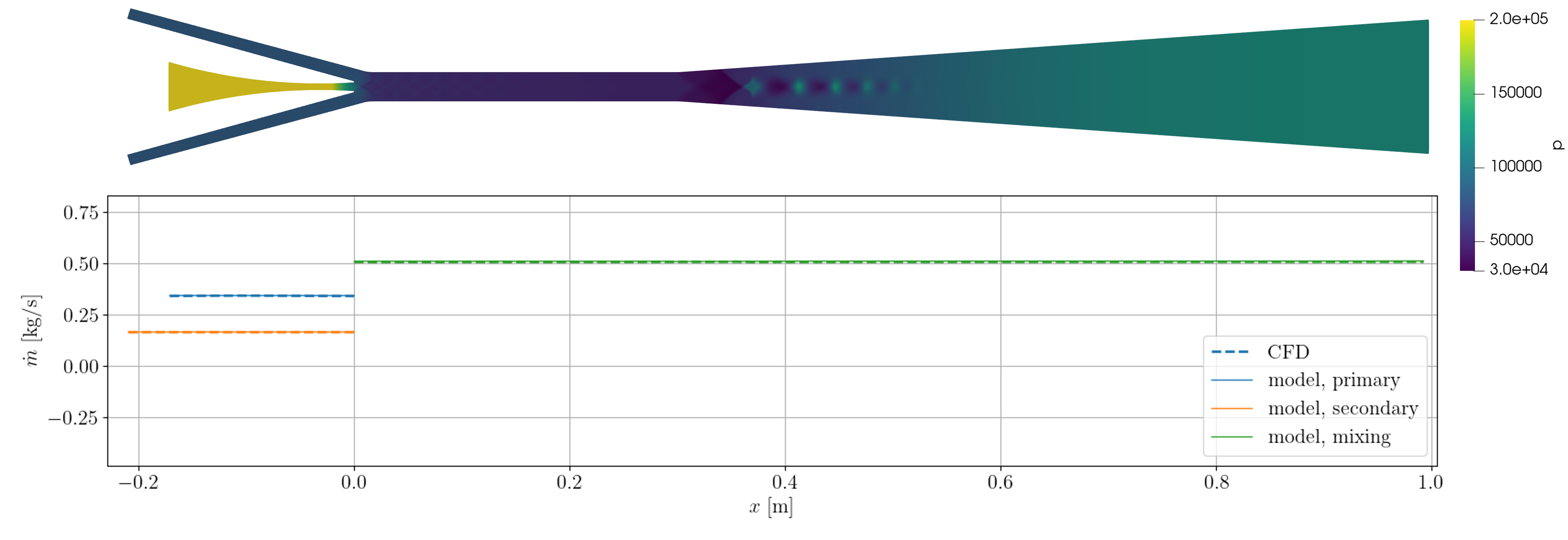}
		\subcaption{$t = 0$ ms}
		\label{fig:CFD_p_mass_0}
	\end{subfigure}
	\\
	\begin{subfigure}[t]{\linewidth}
		\includegraphics[width=\columnwidth]{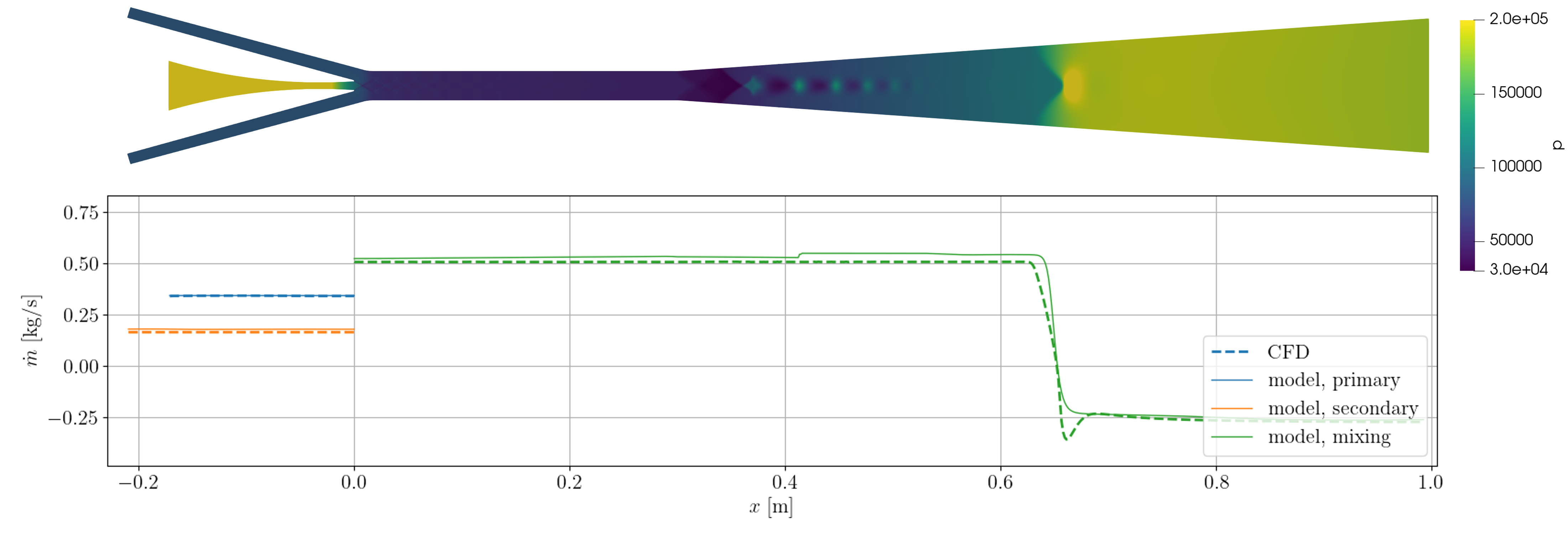}
		\subcaption{$t = 1.0$ ms}
		\label{fig:CFD_p_mass_1.0}
	\end{subfigure}
	\\
	\begin{subfigure}[t]{\linewidth}
		\includegraphics[width=\columnwidth]{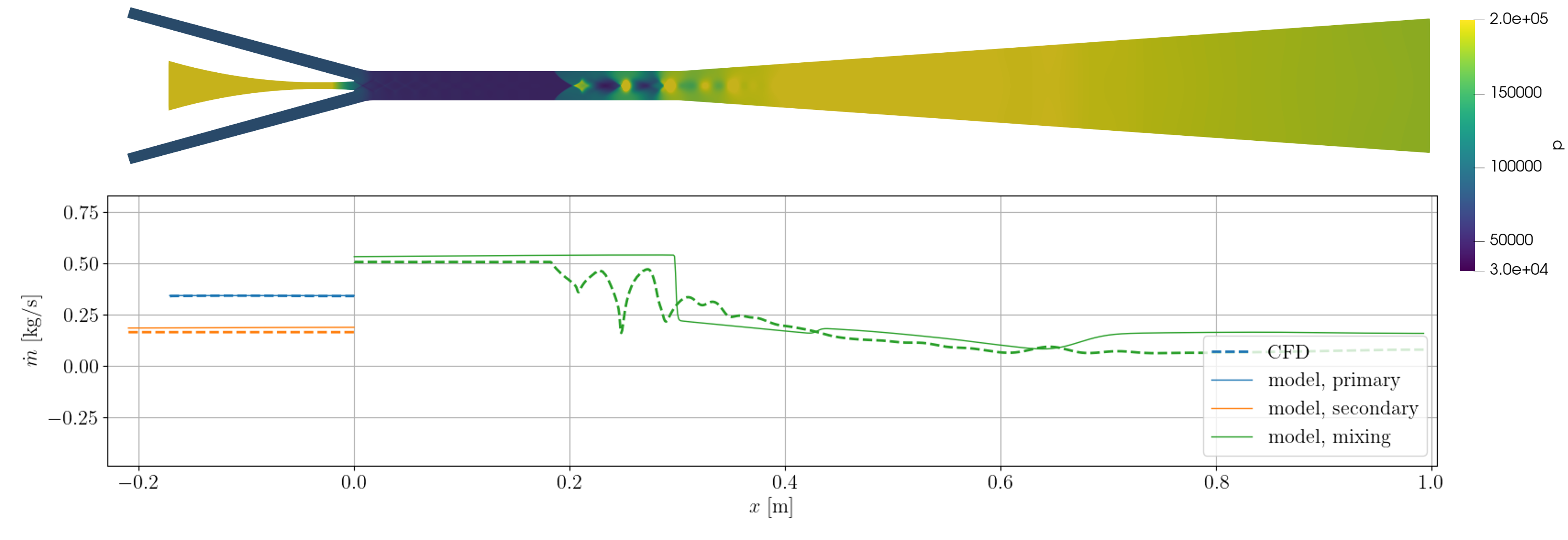}
		\subcaption{$t = 2.4$ ms}
		\label{fig:CFD_p_mass_2.4}
	\end{subfigure}
\end{figure*}

\begin{figure*}[!htb]\ContinuedFloat
    \centering
	\begin{subfigure}[t]{\linewidth}
		\includegraphics[width=\columnwidth]{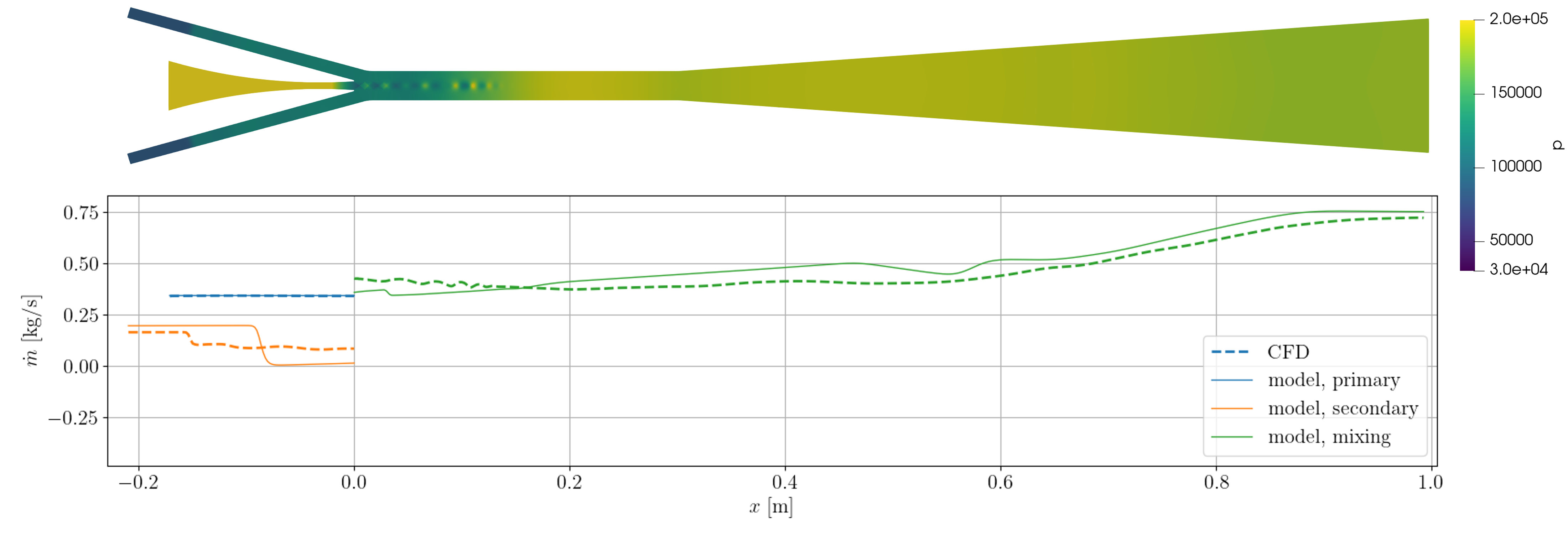}
		\subcaption{$t = 4.4$ ms}
		\label{fig:CFD_p_mass_4.4}
	\end{subfigure}
	\\
	\begin{subfigure}[t]{\linewidth}
		\includegraphics[width=\columnwidth]{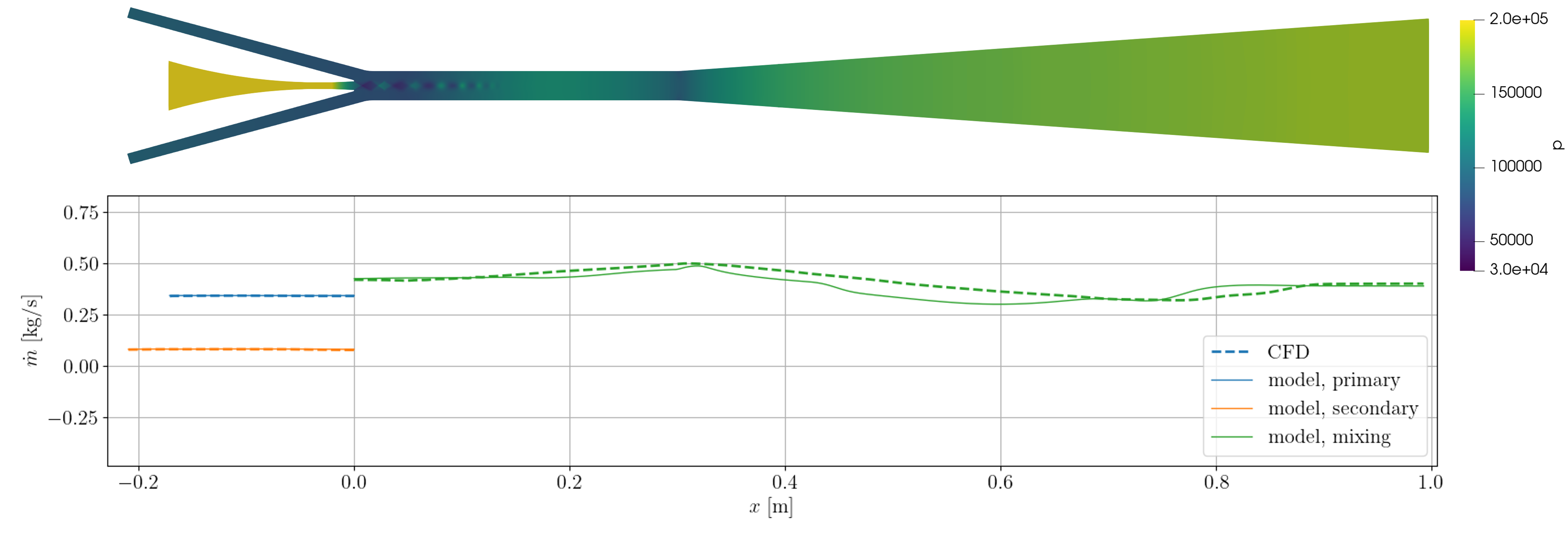}
		\subcaption{$t = 7.0$ ms}
		\label{fig:CFD_p_mass_7.0}
	\end{subfigure}
	\\
	\begin{subfigure}[t]{\linewidth}
		\includegraphics[width=\columnwidth]{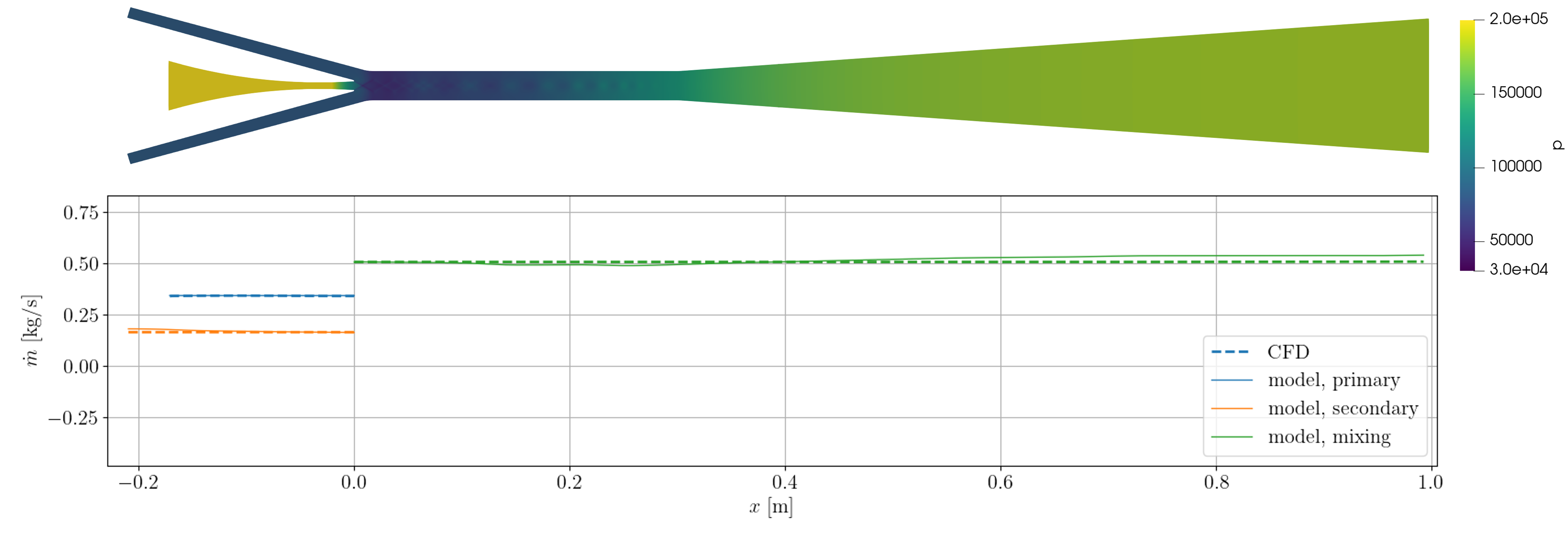}
		\subcaption{$t = 32.4$ ms}
		\label{fig:CFD_p_mass_32.4}
	\end{subfigure}
    \caption{Sequence of snapshots in the transient test case, showing the evolution of the pressure field in the CFD simulation together with the mass flow rates distribution from the CFD and the proposed 1D model. A sudden increase in back pressure from the steady state in Figure \ref{fig:CFD_Mach} creates a shock wave traveling upstream. This makes the temporarily reverse. The shock wave is reflected at the secondary inlet, leading to oscillations that die out in approximately 30 ms, after which the ejector reaches the new steady state. The 1D model captures the transient in mass flow rate and smoothly switches between the ejector's operating modes. The scale of the static pressure is limited to 2 bar for better visualisation. A similar match is obtained for the momentum and total enthalpy flow rates (not shown for conciseness). An animation of these snapshots is included in the supplementary material. }
    \label{fig:CFD_p_mass}
\end{figure*}

\section{Conclusion}\label{sec:6}

We presented a new 1D transient model for ejectors based on a pipeline network analogy. The model combines 1D unsteady Euler equations with a junction model and allows the study of ejectors' performances at a fraction of the cost of classic CFD. The model formulates the 1D gas dynamics for the mixing section in terms of `compound-average' states and describes the stream-wise evolution of integral quantities (e.g. mean velocity and momentum fluxes) and the mixture's thermodynamic conditions (e.g. pressure and Mach number). This formulation allows overcoming the classic quasi-steady assumption. A procedure for automatically calibrating the model on experimental and numerical data is also presented.

To the authors' knowledge, the presented model is the first to allow a low-cost transient simulation of ejectors without resorting to CFD.
The model was successfully validated and benchmarked in steady-state and transient conditions. In the steady state, the model proved capable of correctly predicting the entrainment ratio over a wide range of conditions for which experimental data is available in the literature. In transient conditions, the model proved capable of reproducing travelling waves in the mixing and secondary sections and back-flows due to a sudden change in the operating conditions. 

Future extensions of the work could include releasing the assumption of choked primary flow, enabling the modelling of ab-normal conditions, and formulating closure laws for the model parameters that depend on the operating conditions.

\section*{Acknowledgments}

This work is supported by the European Union’s H2020-EU.3.4 programme, project EJEMOD (Engine bleed JEt pumps continuous behaviour MODelization, grant agreement 101008100). The authors thank Dr. Olivier Lamberts for providing the geometry and the mesh used in the 2D CFD simulations.

\appendix
\section{Fundamental relations}\label{AppA}

Useful relations from elementary gas dynamics are here recalled for self-consistency:

\begin{align}\label{eq:state_quantities}
    e &= e_t - \frac{1}{2} u^2   & T_t & = T \left(1 + \frac{\gamma-1}{2}M^2\right)\nonumber\\
    T &= \frac{e}{C_v}           & p_t & = p \left(1 + \frac{\gamma-1}{2}M^2\right)^\frac{\gamma}{\gamma-1}\nonumber\\
    p &= \rho R T                & h &= e + p/\rho\nonumber\\
    a &= \sqrt{\gamma R T}       & h_t &= e_t + p/\rho\nonumber\\
    M &= \frac{u}{a}             & s &= C_v \ln\left(\frac{p}{\rho^\gamma}\right)
\end{align}

\section{Numerical details of the Roe scheme}\label{AppB}
The Roe scheme provides the fluxes of the Euler equations by solving an approximate Riemann problem at the boundary between each pair of cells. The flux depends on the state $\bm{U}_L$ in the left cell and the right state $\bm{U}_R$. First, the so-called Roe-average state is computed as follows:
\begin{align}
    \tilde{u} &= \frac{\sqrt{\rho_L} u_L + \sqrt{\rho_R} u_R}{\sqrt{\rho_L} + \sqrt{\rho_R}}\\
    \tilde{h}_t &= \frac{\sqrt{\rho_L} h_{t,L} + \sqrt{\rho_R} h_{t,R}}{\sqrt{\rho_L} + \sqrt{\rho_R}}\\
    \tilde{a}^2 &= \left(\gamma-1\right)\left(\tilde{h}_t - \frac{1}{2} \tilde{u}^2\right)
\end{align}
These are used to calculate the eigenvalues $\tilde{\lambda_k}$ and the eigenvectors $\tilde{\bm{\text{K}}}_k$ of the Roe matrix (which approximates the non-linear Jacobian of the Euler equations) and the wave strengths $\tilde{\alpha}_k$:

\begin{align}
    \tilde{\lambda}_1 &= \tilde{u} - \tilde{a}\\
    \tilde{\lambda}_2 &= \tilde{u}\\
    \tilde{\lambda}_3 &= \tilde{u} + \tilde{a}
\end{align}
\begin{align}
    \tilde{\bm{K}_1} &= \left[1, \tilde{u} - \tilde{a}, \tilde{h_t} - \tilde{u} \tilde{a}\right]\\
    \tilde{\bm{K}}_2 &= \left[1, \tilde{u}, \frac{1}{2}\tilde{u}^2\right]\\
    \tilde{\bm{K}}_3 &= \left[1, \tilde{u} + \tilde{a}, \tilde{h_t} + \tilde{u} \tilde{a}\right]
\end{align}
\begin{align}
    \tilde{\alpha}_1 &= \frac{1}{2 \tilde{a}^2}\left(\Delta p - \tilde{a}\tilde{\rho}\Delta u\right)\\
    \tilde{\alpha}_2 &= \Delta \rho - \frac{1}{\tilde{a}^2}\Delta p\\
    \tilde{\alpha}_3 &= \frac{1}{2 \tilde{a}^2}\left(\Delta p + \tilde{a}\tilde{\rho}\Delta u\right)
\end{align} where $\Delta \bullet = \bullet_R - \bullet_L$ acts as a subtracting operator between the right and left state. These quantities allow to calculate the fluxes in equation \eqref{eq:num_scheme} with equation \eqref{eq:Roe_flux}.

\bibliographystyle{elsarticle-num} 
\bibliography{Van_den_Berghe_et_all_2022}


%
%
%
%

\end{document}